# Multiple Cell Upset Cross-Section Uncertainty in Nanoscale Memories: Microdosimetric Approach

G. I. Zebrev, K. S. Zemtsov, R. G. Useinov, M. S. Gorbunov, V. V. Emeliyanov, A. I. Ozerov

*Abstract* —We found that the energy deposition fluctuations in the sensitive volumes may cause multiplicity scatters in the multiple cell upsets in the nanoscale (with feature sizes less than 100 nm) memories.

## I. Introduction

Aggressive scaling of microelectronics components leads to decreased immunity of the digital integrated circuits to external transients due to reducing in noise margin. In particular, commercial highly-scaled digital memories become extremely susceptible to the single event effects (SEE) because of their low critical charges and small sizes [1]. Scaling has spatial and energetic aspects, namely, the dimensional shrinking and the supply voltage reduction. This leads to several important consequences in the context of susceptibility to ionizing particles. First of all, the size of a memory cell turns out to be less than the lateral dimensions of the heavy ion tracks. This problem manifests itself as the multiple cell upsets (MCUs) which are defined as simultaneous errors in more than one memory cell induced by a single particle hit [2]. Secondly, due to both scaling and reduction of the supply voltage, the memory cell critical charge magnitudes $Q_C$ are reducing slowly to the sub-femto-Coulomb region. Such values of the collected charge (of order $10^3 - 10^4$ carriers) correspond to the mean deposited energy as small as a few keVs and average values of critical linear energy transfer (LET) less than 1 MeV-cm$^2$/mg [3]. For instance, the critical charges of the SRAM cells fabricated in a 65 nm SOI technology are reportedly estimated between 0.14 fC (or ~ 880 electrons) and 0.24 fC (or 1500 electrons) [4] corresponding to critical energy 3.2 keV and 5.4 keV, respectively. Energy deposition of such small scales implies a failure of the approaches based entirely on the average dose or average LET concepts. Average LET concept is assumed to be appropriate for only relatively low-scaled ICs (>100 nm), having critical charges > 10 fC [5].

G. I. Zebrev, K. S. Zemtsov, R. G. Useinov, M.S. Gorbunov are with the Department of Micro- and Nanoelectronics of National Research Nuclear University MEPHI, Moscow, Russia, e-mail: gizebrev@mephi.ru.

M.S. Gorbunov is also with Scientific Research Institute of System Analysis, Russian Academy of Sciences (SRISA), Moscow, Russia.

V. V. Emeliyanov, A. I. Ozerov and also R. G. Useinov are with RISI, Lytkarino, Moscow region.

Generally, an average energy deposition at such low magnitudes turns out to be of the same order as energy-loss fluctuations (straggling) [6]. Role of straggling in SEE were discussed also in [7, 8, 9, 10]. The physical reason for importance of straggling in highly-scaled ICs stems from the fact that typical magnitude of energy transfer in elementary interaction between ion and electrons (~ tens of keV ) turns out to be greater than the cell's critical energy.

This means that a soft bit upset can, in principle, be produced by only a single secondary ("delta") electron. Similar effect is likely reported recently in [11], where the "electron-induced SEUs" refer to events in which the initiating particle is a high-energy electron (delta-ray); the eventual upsets are produced by thermalized electron–hole pairs generated as the delta-rays lose their energy through ionization."

For low integration, we have a rather large critical energy $\varepsilon_C$ and an upset occurs if only energy deposition is large enough $\langle \Delta E \rangle > \varepsilon_C$. For highly scaled memories the equality $\langle \Delta E \rangle > \varepsilon_C$ may take place even for extremely low-LET ions such as low-energy proton [12]. High-LET heavy ions are capable to provide a multiple cell upset condition $\langle \Delta E \rangle > n\varepsilon_C$, where *n* are integers up to 10-20. The problem is that energy deposition $\Delta E$ is a stochastic variable, fluctuating due to energy-loss straggling from one ion to another even for the same mean LETs. The main objective of this work is to reveal a role of the energy deposition fluctuations in determining of the MCU multiplicity distribution.

## II. Theoretical background

### A. Local and non-local impact of a single particle

The key device characteristic that determines the upset sensitivity of a device is its critical charge $Q_C$. This charge is defined as the amount of charge that must be released and collected at the terminal of the device to cause the single event effect [13]. It is assumed that any excess energy deposition above a critical value $\varepsilon_C$ in a sensitive volume leads immediately to a single bit upset occurrence. This is a very strong assumption, being essentially microdosimetric one, suggests a tight coupling between the circuit response and microdosimetry of energy deposition within the very small sensitive micro-volumes [14]. Energy deposition is a random variable, and it is especially noticeable on very small spatial scales of modern memory cells. Furthermore, small values of critical energy imply discrete and discontinuous interaction of radiation with the IC material on the small scales of memory cells, which can manifest it in macroscopic properties. A single ionizing ion in highly scaled memories is able to produce a change in macroscopic properties of an ensemble of memory cells (multiple upset). Electron-hole pairs in silicon are produced eventually by energetic secondary electrons (delta electrons) which are coincident in time with the primary ions.

*B. Experimental multiplicity distribution*

Test data are impacted by variations caused by statistics and indeterminacy of the ionization and charge collection processes. Particularly, energy deposition fluctuations, or, the energy-loss straggling, cause substantial indeterminacy in MCU numbers. Indeed, the relative error in number of single bit errors (SBU) can be represented as a sum at least of the two independent terms [15]

$$\frac{\langle \delta N_{SBU}^2 \rangle}{\langle N_{SBU} \rangle^2} = \frac{1}{A_m \Phi} + \frac{\langle \delta n^2 \rangle}{\langle n \rangle^2}, \quad (1)$$

where $\Phi$ is fluence, $A_m\Phi$ is a mean number the ion hits into the memory region square $A_m$. The former term in (1) can be reduced due to good event statistics, while the multiplicity variance $\langle \delta n^2 \rangle$ is controlled by internal mechanisms of energy deposition and charge collection.

We proposed in [15] that the multiplicity variance is likely caused by fluctuations in energy deposition (energy-loss straggling) during the passage of a single ion and cannot be reduced experimentally

$$\frac{\langle \delta n^2 \rangle}{\langle n \rangle^2} = \frac{\Omega_B^2}{\langle \Delta E \rangle^2} = \frac{2}{B}\frac{T_{max}}{\langle \Delta E \rangle} = \frac{\langle T \rangle}{\langle \Delta E \rangle} = \frac{1}{\kappa}, \quad (2)$$

where $\kappa$ is an average number of electron-ion interaction in the sensitive region (see Appendix B).

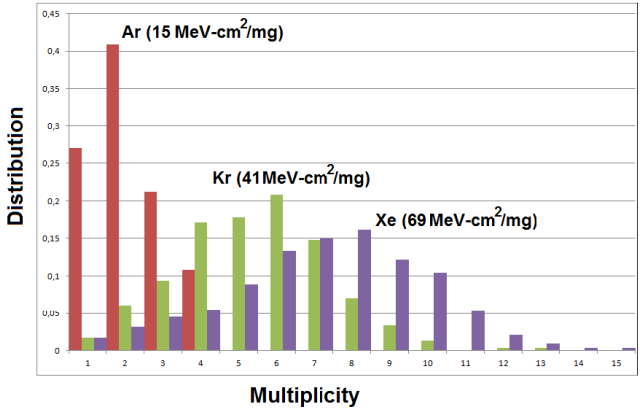

Fig. 1 Multiplicity distributions for IC with technology node 90 nm for different LETs [15]. All ions correspond approximately to the same specific energy ~ 3.5 MeV per a nucleon.

Fig. 1 shows detailed statistical information about multiplicity distributions, obtained by comparing physical and logical upset addresses in the 90 nm memory [15]. The multiplicity distributions for different ions (LETs) are characterized by two remarkable features. First, the mean value of multiplicity of a given ion is approximately proportional to LET. Second, the variances of the multiplicity distributions are also proportional to the LETs. The former evidence shows that mean multiplicity and cross-section are proportional to mean LET (see Appendix D). Respectively, the multiplicity variance for a fixed ion specific energy and different Z turns out also approximately to be proportional to energy deposition variance.

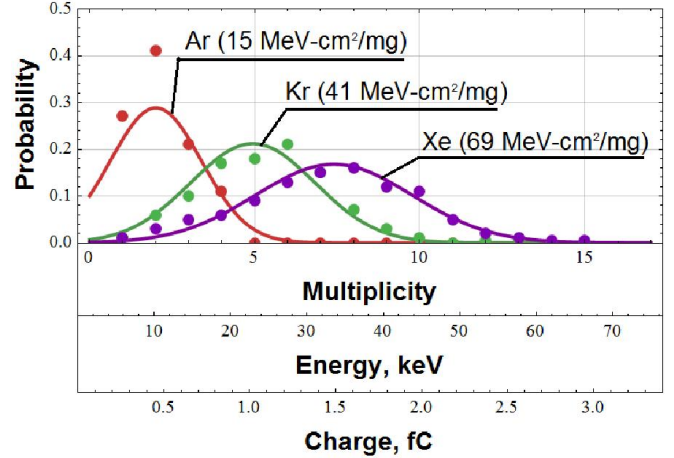

Fig. 2. Comparison distributions for IC with technology node 90 nm for different LETs, nominal layout cell square $a_{cell}$ = 0.8 um$^2$. Extracted parameter $t_{eff}$ = 2.6 nm, $\varepsilon_C$ = 4.5 keV, $Q_C$ = 0.2 fC. $\eta$(Ar) = 1, $\eta$(Kr) = 0.9, $\eta$(Xe) = 0.8.

Figure 2 shows the experimental multiplicity distribution for different ions combined with the analytic distributions of energy-loss straggling (see Appendices B, C ). The values of the critical energy and charge, effective charge collection length are determined by a fitting procedure. Interestingly, the extracted value of the critical charge corresponds almost exactly to an estimation with an old empirical formula [16]
$Q_C \cong 23(L_{node}/\mu m)^2$ fC ~ 0.16 fC for $L_{node}$ = 90 nm.

*C. MCU cross-section model*

The main conjecture of the proposed model is to assume that the critical energy of n-fold upset equals to $n\varepsilon_C$, provided that the critical energy of a single failure is $\varepsilon_C$. The introduction of the critical charge concept for memory cell $\varepsilon_C$ implies, in fact, a use of the microdosimetric approach to the calculation of the upset cross-sections and rates. It was shown in [8, 9] that soft error rate (SER) can be represented as an average over the distribution of the chords lengths in the sensitive volume and over the LET spectrum of space environment. This approach is equivalent to the CREME96 ideology, which, in essence, is also based on the microdosimetric concept.

Following a general approach proposed in [9] one can represent the dependence of mean MCU cross-section on LET as a superposition of the step response functions averaged over energy-loss straggling distribution (see Appendix D)

$$\langle \sigma(\Lambda) \rangle \cong \frac{a_{cell}}{2} \sum_{n=1}^{\max n} \mathrm{erfc}\left(\frac{n\varepsilon_C - \Lambda t_{eff}}{\sqrt{2n}\,\Omega_{B0}}\right) \quad (3)$$

Physical meaning of this equation is obvious from its visualization in Fig. 3 illustrating a staircase-like view of the cross-section vs LET dependence. A condition $n\varepsilon_C < \Delta E < (n+1)\varepsilon_C$ corresponds to the multiple cell upset case with the multiplicity equal to $n$.

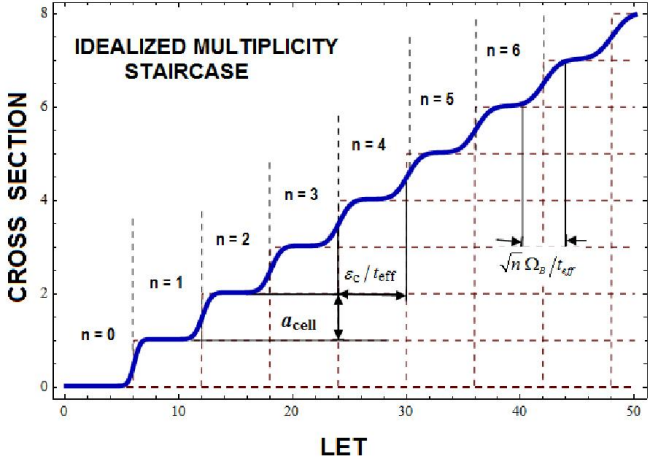

Fig. 3. Cross-section dependence illustrating an ideal view staircase-like cross-section dependence plotted for a case of single critical charge and low dispersion of energy deposition and other scatter mechanisms. Cross-section is normalized by the cell area $a_{cell}$.

Fig. 3 shows an idealized case $\Omega_B \ll \varepsilon_C$ when the energy deposition variance $\Omega_B$ is much less than the cell critical energy. In practice, a staircase view cross-section vs LET dependence (a priori presumed earlier [17, 18]) is not observed due to presence of the statistical errors and variety of random factors.

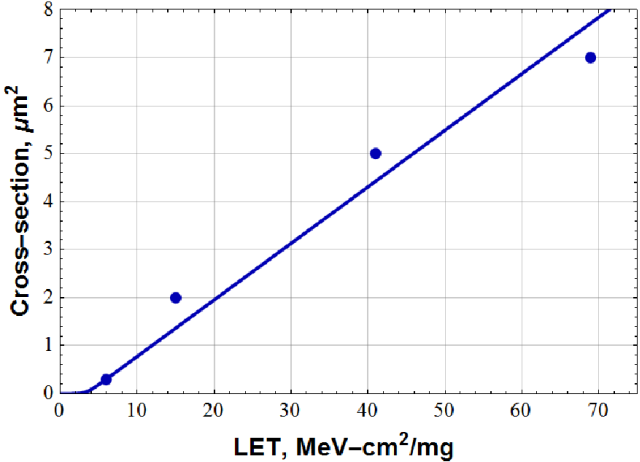

Fig. 4. MCU cross-section computed as function of independent parameter in Eq.# with $a_{cell} = 0.8$ um$^2$ (layout parameter), $t_{eff} = 2.6$ nm, $\varepsilon_C = 4.1$ keV, $Q_C = 0.2$ fC, calculated with specific energy 3 MeV/u, $\eta = 0.9$.

In particular, Fig. 4 shows comparison of the simulated (with Eq. 3) curve and experimental points of experimental dependence for mean cross-section obtained in independent experiment for the same memory IC as in Fig. 1 [15]. Thus, a quasi-linear cross-section dependence emerges here as a staircase, completely smoothed by the energy-loss fluctuations. The average slope of the smoothed cross-section vs LET dependence $K_d$ [15] can be estimated as follows.

$$\langle \sigma \rangle = a_{cell} \langle n \rangle = a_{cell} \eta \frac{\Lambda t_{eff}}{\varepsilon_C} \rho_{Si} \qquad (4)$$

### D. Restrictions of the model

A simplified form of proposed model is implicitly based on a number of simplifying assumptions: (a) there is a single value of critical charge and error types, implying a uniform view of idealized cross-section staircase; (b) it is assumed that energy deposition and charge collection occur in an uncorrelated way. In fact, these processes may be slightly or strongly correlated due to the constraints in the charge collection, imposed by the circuit architecture and layout.

### III. CONCLUSION

We present experimental evidence that the MCU average cross-section and multiplicity and its variations are determined respectively by the average energy deposition and energy-loss straggling.

### IV. APPENDIX

#### A. Average MCU cross-section via multiplicity

A partial cross-section of the $n$-fold MCU can be defined as proportional to the number of the $n$-fold upset [15]. The distribution partial cross-sections over multiplicities conforms to the corresponding probability distribution $p_n$ defined as

$$p_n(\Lambda) = \frac{\sigma_n(\Lambda)}{a_{cell}}. \qquad (A1)$$

where $a_{cell}$ is elementary memory cell area. The average multiplicity $\langle n(\Lambda) \rangle = \sum n\, p_n(\Lambda)$ and average cross-section $\langle \sigma(\Lambda) \rangle = \sum n\, \sigma_n(\Lambda)$ turn out thusly to be coupled as follows [15]

$$\langle \sigma(\Lambda) \rangle = a_{cell} \langle n(\Lambda) \rangle. \qquad (A2)$$

This simple relationship reveals a deep connection between the mean MCU cross-section and multiplicity. It means that normalizing the experimental dependence $\langle \sigma(\Lambda) \rangle$ by the characteristic area of the memory layout $a_{cell}$ one can easily gets the dependence of the mean multiplicity as function of incoming ion's LET.

#### B. Energy-loss straggling and distribution

Every heavy ion, passing through the sensitive volume, loses different amounts of energy due to stochastic and discrete character of interaction with the material electrons. Energy deposition fluctuations (energy-loss straggling) can be characterized approximately by the Bohr distribution [6]

$$P_B(\Delta E, \langle \Delta E \rangle) = \frac{1}{\sqrt{2\pi}\Omega_B} \exp\left(-\frac{(\Delta E - \langle \Delta E \rangle)^2}{2\Omega_B^2}\right), \qquad (B1)$$

where $\langle \Delta E \rangle$ is the average energy deposition, which can be calculated (in keVs) as

$$\langle \Delta E \rangle = \Lambda\, \rho_{Si}\, t_{eff} \cong 0.232\, \Lambda t_{eff} \text{ keV}, \qquad (B2)$$

where $\Lambda$ is the LET (in MeV×cm$^2$/mg), $\rho_{Si}$ is the Si mass density in g/cm$^3$, $t_{eff}$ is the effective collection length (in nm).

The variance of energy deposition a non-relativistic case may be represented as follows

$$\Omega_B^2 = (2/B) T_{\max} \langle \Delta E \rangle, \quad \text{(B3)}$$

where $B = \ln(T_{\max}/I_{ion})$ is the Coulomb logarithm, $I_{ion}$ is the ionization energy of the material (~ 0.173 keV in Si). The maximum energy $T_{max}$ that can be transferred from the projectile ion to an electron in the silicon via the Coulomb interaction mechanism is given in non-relativistic case by the formula

$$T_{\max} = 4 \frac{m_e}{M_{ion}} E_{ion} = 4 \frac{m_e c^2}{E_0} \varepsilon \cong \frac{\varepsilon}{456}, \quad \text{(B4)}$$

where $m_e$ is the free electron mass, $M_{ion}$ and $E_{ion}$ are the ion mass and energy respectively, $\varepsilon$ is the ion specific energy in MeV/nucleon.

### C. Multiplicity straggling

Eq.B1 can be rewritten down as the multiplicity distribution around the mean value $\langle n \rangle$

$$P_B(n, \langle n \rangle) = \frac{1}{\sqrt{2\pi \langle n \rangle / m}} \exp\left(-\frac{(n - \langle n \rangle)^2}{2 \langle n \rangle / m}\right), \quad \text{(C1)}$$

where the mean multiplicity $\langle n \rangle$ and the cell's critical energy $\varepsilon_C$ are connected as follows

$$\langle n \rangle \varepsilon_C = \eta \langle \Delta E \rangle \quad \text{(C2)}$$

where $\eta$ ($\leq 1$) is the charge yield efficiency, $m$ is mean number of Coulomb interactions between the ion and medium electrons required for a single bit upset

$$m = \frac{1}{\eta} \frac{B}{2} \frac{\varepsilon_C}{T_{\max}} = \frac{\varepsilon_C}{\eta \langle T \rangle} \quad \text{(C3)}$$

$\langle T \rangle = 2 T_{\max} / B$ is an average energy transfer per an electron-ion interaction. Notice that the multiplicity variance calculated by averaging with Eq.C1 $\langle \delta n^2 \rangle = \langle (n - \langle n \rangle)^2 \rangle = \langle n \rangle / m$ is in full accordance with Eq.2 where $\kappa = m \langle n \rangle$.

### D. Staircase cross-section

Following a general approach proposed in [9] the cross-section of the *n*-fold upset is proportional an averaged value of the step response function $\langle \theta(n \varepsilon_C - \Delta E) \rangle_{AV}$, where $\langle ... \rangle_{AV}$ is defined as averaging over the chord length distribution, LET spectrum and other variation factors. Particularly, even for a fixed LET and chord length, still remains variations due to the energy-loss straggling. Then, a mean cross-section for a given LET can be obtained by averaging over energy deposition fluctuations $\Delta \varepsilon = \Delta E - \langle \Delta E \rangle = \Delta E - \Lambda t_{\mathit{eff}}$ around its mean value

$$\langle \sigma(\Lambda) \rangle = a_{cell} \sum_{n=1} \langle \theta(n \varepsilon_C - \Delta E) \rangle = a_{cell} \sum_{n=1} \langle \theta(n \varepsilon_C - \Lambda t_{\mathit{eff}} + \Delta \varepsilon) \rangle \quad \text{(D1)}$$

Using Eq.B1, one can obtains

$$\langle \theta(n \varepsilon_C - \Delta E) \rangle = \int_{-\infty}^{\infty} P(\Delta E, n \varepsilon_C) \theta(n \varepsilon_C - \Delta E) d(\Delta \varepsilon) \cong$$

$$\cong \frac{1}{2} \operatorname{erfc}\left( \frac{n \varepsilon_C - \Lambda t_{\mathit{eff}}}{\sqrt{2n} \, \Omega_{B0}} \right) \quad \text{(D2)}$$

$$\Omega_{B0}^2 = (2/B) T_{\max} \varepsilon_C \quad \text{(D3)}$$

where erfc(*x*) is the complementary error function.